\documentstyle[amstex]{article}

\newlength{\defaultparindent}
\newenvironment{Default Paragraph Font}{}{}
\input{tcilatex}

\begin{document}

\begin{center}
{\Large An alternative method for calculating the energy of gravitational
waves}

Miroslav S\'{u}ken\'{i}k and Jozef Sima

Slovak Technical University, Radlinsk\'{e}ho 9, SK-812 37 Bratislava,
Slovakia
\end{center}

Abstract

In the expansive nondecelerative universe model, creation of matter occurs
due to which the Vaidya metrics is applied. This fact allows for localizing
gravitational energy and calculating the energy of gravitational waves using
an approach alternative to the well established procedure based on
quadrupole formula. Rationalization of the gradual increase in entropy of
the Universe using relation describing the total curvature of space-time is
given too.

As a source of gravitational waves, any physical system with time dependent
mass distribution can be considered. The amount of such energy, emitted
within a time unit is described by known quadrupole formula:

$\frac{dE}{dt}=P_{qw}=-\frac{G}{45c^{5}}\dddot{K}_{\alpha \beta }\qquad \ \
\ \ \ \ \ \ \ \ \ \ \ \ \ \ \ \ \ \ \ \ \ \ \ \ \ \ \ \ \ \ \ \ \ \ \ \ \ \
\ \ \ \ \ \ \ \ \ \ \ \ \ \ \ \ \ \ \ \ \ \ \ \ \ \ \ \ \ \ \ \ \ \ \ \ \ \
\ \ \ \ \ \ \ \ $(1)

where $K_{\alpha \beta }$ is the tensor of quadrupole mass distribution in
the source of emission. For $K_{\alpha \beta }$ it holds:

$K_{\alpha \beta (t)}=\int \rho _{(t,x)}.(3X_{\alpha }X_{\beta }-\delta
_{\alpha \beta }X_{\kappa }X^{\kappa })\Delta V$ \qquad\ \ \ \ \ \ \ \ \ \ \
\ \ \ \ \ \ \ \ \ \ \ \ \ \ \ \ \ \ \ \ \ \ \ \ \ \ \ \ \ \ \ \ \ \ \ \ \ \
\ \ \ \ \ \ \ \ \ \ (2)

In the expansive nondecelerative universe (ENU) model [1], the creation of
matter and of gravitational energy simultaneously occur. The laws of energy
conservation still hold since the energy of gravitational field is negative
in ENU. The total energy of the Universe is thus exactly zero [2] and the
Universe can continuously expand with the velocity of light $c$. This
postulate is expressed in the ENU by equation:

$a=c.t_{c}=\frac{2GM_{u}}{c^{2}}\qquad \ \ \ \ \ \ \ \ \ \ \ \ \ \ \ \ \ \ \
\ \ \ \ \ \ \ \ \ \ \ \ \ \ \ \ \ \ \ \ \ \ \ \ \ \ \ \ \ \ \ \ \ \ \ \ \ \
\ \ \ \ \ \ \ \ \ \ \ \ \ \ \ \ \ \ \ \ \ \ \ \ \ \ \ \ \ \ \ \ \ \ \ \ \ \
\ $(3)

where a is the gauge factor, $t_{c}$ is the cosmological time, $M_{u}$ is
the mass of ENU.

In such a model, due to the matter creation the Vaidya metrics [3] must be
used which enables to localise gravitational energy. Weak fields are in
first approximation described by Tolman's relation [4]

$\epsilon _{g}=-\frac{R.c^{4}}{8\pi .G}=-\frac{3m.c^{2}}{4\pi a.r^{2}}\qquad
\ \ \ \ \ \ \ \ \ \ \ \ \ \ \ \ \ \ \ \ \ \ \ \ \ \ \ \ \ \ \ \ \ \ \ \ \ \
\ \ \ \ \ \ \ \ \ \ \ \ \ \ \ \ \ \ \ \ \ \ \ \ \ \ \ \ \ \ \ \ \ \ \ \ \ \
\ \ \ \ \ \ \ \ \ \ \ \ $(4)

in which $\varepsilon _{g}$ is the density of the gravitational energy being
emitted by a body with the mass $m$ at the distance $r$, $R$ denotes the
scalar curvature (contrary to the more frequently used Schwarzschild
metrics, in the Vaidya metrics,

$R\neq $ 0).

Using relations (3) and (4), equation expressing gravitational output $P_{g}$
is obtained:

$P_{g}=\frac{d}{dt}\int \epsilon _{g}dV=-\frac{m.c^{3}}{a}=-\frac{m.c^{2}}{%
t_{c}}$ \qquad\ \ \ \ \ \ \ \ \ \ \ \ \ \ \ \ \ \ \ \ \ \ \ \ \ \ \ \ \ \ \
\ \ \ \ \ \ \ \ \ \ \ \ \ \ \ \ \ \ \ \ \ \ \ \ \ \ \ \ \ \ \ \ \ \ \ \ \ (5)

If the total mass of universe, $M_{u}$ is substituted from (3) to (5),
relation for the total output of the gravitational energy $P_{tot}$ (its
absolute value) by the Universe appears

$P_{tot}=-\frac{c^{5}}{2G}\approx \left| 2\times 10^{52}\right| W\qquad \ \
\ \ \ \ \ \ \ \ \ \ \ \ \ \ \ \ \ \ \ \ \ \ \ \ \ \ \ \ \ \ \ \ \ \ \ \ \ \
\ \ \ \ \ \ \ \ \ \ \ \ \ \ \ \ \ \ \ \ \ \ \ \ \ \ \ \ \ \ \ \ \ \ \ \ \ \
\ \ $(6)

This output is time independent and represents at the same time the upper
limit of the gravitational energy output. No machine or physical phenomenon
may create a higher output than that of $P_{tot}$. To illustrate the meaning
of the above statement, radiant output of our galaxy is about 10$^{37}$ W,
number of galaxies in the perceptible part of the Universe is about 10$^{11}$
which corresponds to the radiant output amounting 10$^{48}$ W.

Sources of the gravitational waves can be both periodical and aperiodical.
As an example of aperiodical sources, accelerated direct motion, burst of
supernova or nonspheric gravitational breakdown can serve. Further we will
focus our attention to periodical sources such as planets rotating around
the Sun or a double star rotating around the common centre of inertia. The
emitted gravitational energy cannot exceed $P_{tot}$ value from (6). In any
case, a condition (7) must be observed

$P_{gw}\leq E_{k}$.$\omega P\leq \frac{c^{5}}{2G}\qquad \ \ \ \ \ \ \ \ \ \
\ \ \ \ \ \ \ \ \ \ \ \ \ \ \ \ \ \ \ \ \ \ \ \ \ \ \ \ \ \ \ \ \ \ \ \ \ \
\ \ \ \ \ \ \ \ \ \ \ \ \ \ \ \ \ \ \ \ \ \ \ \ \ \ \ \ \ \ \ \ \ \ \ \ \ \
\ \ \ \ $(7)

where $P_{gw}$ is the energy of gravitational waves emitted within a time
unit, $E_{k}$ means the kinetic energy of a body moving on circular or
elliptic orbit with the angular velocity $\omega $. In these circumstances
(the excentricity of an elliptic orbit is omitted in deriving the following
relations), the ratio of the emitted gravitational energy $P_{gw}$ and the
kinetic energy $E_{k}$ of such body will be comparable to that of $E_{k}$
and $P_{tot}$ (in the limiting case, both value are identical and equal to
unity), i.e.

$\frac{P_{gw}}{E_{k}.\omega }\approx \frac{E_{k}.\omega }{P_{tot}}\qquad \ \
\ \ \ \ \ \ \ \ \ \ \ \ \ \ \ \ \ \ \ \ \ \ \ \ \ \ \ \ \ \ \ \ \ \ \ \ \ \
\ \ \ \ \ \ \ \ \ \ \ \ \ \ \ \ \ \ \ \ \ \ \ \ \ \ \ \ \ \ \ \ \ \ \ \ \ \
\ \ \ \ \ \ \ \ \ \ \ \ \ \ \ \ \ \ \ \ \ \ \ \ \ $(8)

It follows from (8) that the output of emitted gravitational waves is

$P_{gw}=-\frac{2G.E_{k}^{2}.\omega ^{2}}{c^{5}}\qquad \ \ \ \ \ \ \ \ \ \ \
\ \ \ \ \ \ \ \ \ \ \ \ \ \ \ \ \ \ \ \ \ \ \ \ \ \ \ \ \ \ \ \ \ \ \ \ \ \
\ \ \ \ \ \ \ \ \ \ \ \ \ \ \ \ \ \ \ \ \ \ \ \ \ \ \ \ \ \ \ \ \ \ \ \ \ \
\ \ \ \ \ \ \ \ \ $(9)

The validity of equation (9) was tested on a system consisting of two bodies
with the nearly identical masses

$m_{1}\doteq $ $m_{2}$ = $m$ \ \ \ \ \ \ \ \ \ \ \ \ \ \ \ \ \ \ \ \ \ \ \ \
\ \ \ \ \ \ \ \ \ \ \ \ \ \ \ \ \ \ \ \ \ \ \ \ \ \ \ \ \ \ \ \ \ \ \ \ \ \
\ \ \ \ \ \ \ \ \ \ \ \ \ \ \ \ \ \ \ \ \ \ \ \ \ \ \ \ \ \ \ \ \ \ \ \ \ \
\ \ \ \ \ \ (10)

These bodies rotate around the common centre of inertia on circular orbit
with diameter $r$ and angular velocity $\omega $. The kinetic energy of
either of the bodies is

$E_{k}=\frac{1}{2}m.r^{2}.\omega ^{2}$ \ \ \ \ \ \ \ \ \ \ \ \ \ \ \ \ \ \ \
\ \ \ \ \ \ \ \ \ \ \ \ \ \ \ \ \ \ \ \ \ \ \ \ \ \ \ \ \ \ \ \ \ \ \ \ \ \
\ \ \ \ \ \ \ \ \ \ \ \ \ \ \ \ \ \ \ \ \ \ \ \ \ \ \ \ \ \ \ \ \ \ \ \ \ \
\ \ \ \ \ \ \ (11)

Taking the identical masses of the bodies into account, both bodies must
emit the same quantity of gravitational energy. Then it follows from (9),
(10) and (11)

$P_{gw}=-\frac{4G.E_{k}^{2}.\omega ^{2}}{c^{5}}=-\frac{G.m^{2}.r^{4}.\omega
^{6}}{c^{5}}$ \qquad\ \ \ \ \ \ \ \ \ \ \ \ \ \ \ \ \ \ \ \ \ \ \ \ \ \ \ \
\ \ \ \ \ \ \ \ \ \ \ \ \ \ \ \ \ \ \ \ \ \ \ \ \ \ \ \ \ \ \ \ \ \ \ \ \ \
\ \ \ \ \ \ (12)

Quadrupole formula leads to equation

$P_{gw}=-\frac{32G.r^{4}.\omega ^{6}}{5c^{5}}\left( \frac{m_{1}.m_{2}}{%
m_{1}+m_{2}}\right) ^{2}$ \qquad\ \ \ \ \ \ \ \ \ \ \ \ \ \ \ \ \ \ \ \ \ \
\ \ \ \ \ \ \ \ \ \ \ \ \ \ \ \ \ \ \ \ \ \ \ \ \ \ \ \ \ \ \ \ \ \ \ \ \ \
\ \ \ \ \ \ \ \ \ \ \ \ \ \ \ \ (13)

which can be, for cases where (10) and (13) hold, transformed into

$P_{gw}=-\frac{8G.m^{2}.r^{4}.\omega ^{6}}{5c^{5}}\qquad \ \ \ \ \ \ \ \ \ \
\ \ \ \ \ \ \ \ \ \ \ \ \ \ \ \ \ \ \ \ \ \ \ \ \ \ \ \ \ \ \ \ \ \ \ \ \ \
\ \ \ \ \ \ \ \ \ \ \ \ \ \ \ \ \ \ \ \ \ \ \ \ \ \ \ \ \ \ \ \ \ \ \ \ \ \
\ \ \ \ \ $(14)

Equation (14) differs from our relation (12) only in coefficient 8/5 which
can be, bearing in mind the simplifications used in our derivation, assessed
as an excellent agreement.

Further domain of application of our approach lies in possibility to
rationalize some questions which are still open, the entropy of the Universe
in its begining being one of them. A problem arises [5] in interpretation of
the individual elements in a simplified equation expressing the total
curvature of space-time

$R_{\bullet \bullet \bullet \bullet }=W_{\bullet \bullet \bullet \bullet
}+R_{\bullet \bullet }^{o}g_{\bullet \bullet }\qquad \ \ \ \ \ \ \ \ \ \ \ \
\ \ \ \ \ \ \ \ \ \ \ \ \ \ \ \ \ \ \ \ \ \ \ \ \ \ \ \ \ \ \ \ \ \ \ \ \ \
\ \ \ \ \ \ \ \ \ \ \ \ \ \ \ \ \ \ \ \ \ \ \ \ \ \ \ \ \ \ \ \ \ \ \ \ \ \ $%
(15)

where $R$ is the Riemann tensor representing the total curvature of
space-time, $W$ is the Weyl's tensor describing deformation and slap forces, 
$R^{o}$ is the Ricci's tensor, $g$ is the metrics tensor.

Application of the Vaidya metrics manifests (4) that the scalar curvature
decreases in time. Since this curvature is a reduction of the Ricci's
tensor, the latter must decrease in time too. At identical conditions the
Reimann's tensor is time independent which leads to conclusion stating that
the Weyl's tensor must be gradually increasing in time and, in turn, the
Universe had to start its expansion being in a highly ordered state, i.e. in
the state with a minimal entropy [5].

Conclusions

The values of the energy of gravitational waves obtained using our
alternative simple approach based on first approximation applied to the
domain of weak gravitational fields are comparable to those derived by exact
quadrupole relation. In addition to our previous results, the
prospectiveness of the ENU model and Vaidya metrics is newly underlined.

References

[1] \ V. Skalsk\'{y}, M. S\'{u}ken\'{i}k: Astrophys. Space Sci.,178 (1991)
169

[2] \ S. Hawking: Sci. Amer., 236 (1980) 34

[3] \ P.C. Vaidya: Proc. Indian Acad. Sci., A33 (1951) 264

[4] \ J. Sima, M. S\'{u}ken\'{i}k: General Relativity and
Quantum Cosmology, Preprint in: US National Science Foundation,

\ \ \ \ E-print archive: gr-qc\@xxx.lanl.gov., paper 9903090

[5] \ R. Penrose: The Large, the Small and the Human Mind,Cambridge
University Press, 1997, p. 24

\end{document}